\documentclass[prl,aps,twocolumn,superscriptaddress, showpacs]{revtex4-1}
\usepackage{graphicx}
\usepackage{dcolumn}
\usepackage{bm}
\usepackage{color}
\usepackage{bigstrut}
\usepackage{tabularx}
\usepackage{float}
\usepackage{ulem}
\usepackage[]{color}
\usepackage{amsmath}
\newcolumntype{L}[1]{>{\raggedright\arraybackslash}p{#1}}
\newcolumntype{C}[1]{>{\centering\arraybackslash}p{#1}}
\newcolumntype{R}[1]{>{\raggedleft\arraybackslash}p{#1}}

\begin{document}

\title{Observation of compact ferrimagnetic skyrmions in DyCo$_3$ film}

\author{K.~Chen}
\altaffiliation{Email: kaichen.hzg@gmail.com}
  \affiliation{Helmholtz-Zentrum Berlin f\"{u}r Materialien und Energie, Albert-Einstein-Str.15, 12489 Berlin, Germany}
\author{D.~Lott}
  \affiliation{Institute for Materials Research, Helmholtz-Zentrum Geesthacht, 21502 Geesthacht, Germany}
\author{A. Philippi-Kobs}
\affiliation{Deutsches Elektronen-Synchrotron DESY, Notkestraße 85, 22607 Hamburg, Germany}
\author{M.~Weigand}
   \affiliation{Helmholtz-Zentrum Berlin f\"{u}r Materialien und Energie, Albert-Einstein-Str.15, 12489 Berlin, Germany}
  \affiliation{Max-Planck-Institut für Intelligente Systeme, 70569 Stuttgart, Germany} 
\author{C.~Luo}
  \affiliation{Helmholtz-Zentrum Berlin f\"{u}r Materialien und Energie, Albert-Einstein-Str.15, 12489 Berlin, Germany}  
\author{F.~Radu}
  \affiliation{Helmholtz-Zentrum Berlin f\"{u}r Materialien und Energie, Albert-Einstein-Str.15, 12489 Berlin, Germany}  

%\date{\today}
\begin{abstract}
Owing to the experimental discovery of magnetic skyrmions stabilized by the Dzyaloshinskii-Moriya and/or dipolar interactions in thin films, there is a recent upsurge of interest in magnetic skyrmions with antiferromagnetic spins in order to overcome the fundamental limitations inherent with skyrmions in ferromagnetic materials. Here, we report on the observation of compact ferrimagnetic skyrmions for the class of amorphous alloys consisting of 4f rare-earth and 3d transition-metal elements with perpendicular magnetic anisotropy, using a DyCo$_3$ film, that are identified by combining x-ray magnetic scattering, scanning transmission x-ray microscopy, and Hall transport technique. These skyrmions, with antiparallel aligned Dy and Co magnetic moments and a characteristic core radius of about 40~nm, are formed during the nucleation and annihilation of the magnetic maze-like domain pattern exhibiting a topological Hall effect contribution. Our findings provide a promising route for fundamental research in the field of ferrimagnetic/antiferromagnetic spintronics towards practical applications. 
\end{abstract}

\maketitle

Stabilized by the bulk Dzyaloshinskii-Moriya interaction (DMI), skyrmion lattices with swirling magnetic textures have been originally predicted by Bogdanov et al. \cite{Bogdanov2001, Rossler2006} and later observed in non-centrosymmetric single crystal helimagnets \cite{Muhlbauer2009, Yu2010, Jonietz2010, Yu2011, Yu2012, Seki2012, Braun2012, Nagaosa2013, Lin2013, Wiesendanger2016, Fert2017}. The equilibrium conditions~\cite{Jalil2016, Wang2018}, which satisfy the stability criteria for the formation of a robust Bloch-skyrmions lattice, fill only a narrow pocket of the magnetic phase diagram, usually of few tens of milli-Tesla and few degrees Celsius around the onset of the ordering temperature which is well below room temperature. Different approaches have been pursued to either stretch the skyrmion pocket towards room temperature, e.g., by reducing the dimensionality of the crystal using epitaxial thin films~\cite{Huang2012,Porter2014} or by making use of interfacial DMI present in ultrathin films and multilayers leading to the formation of N\'{e}el skyrmions \cite{Fert2013, Iwasaki2013, Moreau-Luchaire2015,Jiang2015, Boulle2016,Woo2016,Woo2017, Litzius2017, Hrabec2017, Soumyanarayanan2017}. For the latter case, recent observations revealed that skyrmions can be created and manipulated at room temperature, which makes such topological protected states particularly exciting for future spintronic applications~\cite{Jonietz2010,Yu2012, Fert2013,Yang2018,zhou2018,back2020}. As exciting, it is suggested that skyrmions can occur also in systems which do not exhibit an intrinsic DMI, but a so called {\it effective} DMI that may result from mechanisms involving magnetic frustration like curvature-induced DMI \cite{Gaididei2014,Sheka2015,Kravchuk2016,Volkov2019}, noncolinear type of magnetic interactions~\cite{Vistoli2019}, and stray fields for systems with relatively strong perpendicular magnetic anisotropy~\cite{Buttner2018}. 

In practice, however, the relative slow and pinning-dominated current-driven dynamic behaviors of the ferromagnetic skyrmions \cite{Jiang2015, Woo2016, Hrabec2017} will limit their potential for future spintronic applications. Moreover, the interaction between spin-polarized currents and ferromagnetic skyrmions result in a non-collinear skyrmion movement in respect to the current flow direction known as the skyrmion Hall effect (SkHE) \cite{Nagaosa2013, Litzius2017, Jiang2017}. The skyrmion Hall effect, which is detrimental for applications, is proportional to the net magnetization. Hence, antiferromagnetic (AF) skyrmions were proposed to suppress the SkHE \cite{Barker2016, Zhang2016}, however, such skyrmions have been experimentally realized so far only in synthetic antiferromagnets~\cite{Legrand2020, Dohi2019, Chen2020}. Ferrimagnetic systems may open an alternative route towards the realization of skyrmion systems suitable for applications. Particularly, ferrimagnetic films of amorphous alloys consisting of 4f rare-earth and 3d transition-metal elements (RE-TM alloys) exhibiting perpendicular magnetic anisotropy are attracting very much attention because of their versatile magnetic properties~\cite{Radu2012, Finazzi2013, Mangin2014, IR2015, Chen2015, Seifert2017, Radu2018, Chen2019}. First realizations of skyrmions in ferrimagnetic RE-TM alloys were reported only recently, e.g., showing a promising reduction of the SkHE \cite{Woo2018} or even a complete suppression of the SkHE at the angular momentum compensation temperature~\cite{Hirata2019}. 

Ferrimagnetic RE-TM alloy films with perpendicular magnetic anisotropy \cite{Chen2015,Chen2015b,Kaneyoshi2018} are generally expected to host an eventual skyrmion state due to the bulk DMI resulting from the asymmetric distribution of the elemental content \cite{Kim2019}, or the weak interfacial DMI induced by the capping layer\cite{Streubel2018}. Notably, the absence of a center of inversion symmetry in amorphous materials inherently causes a non-vanishing DMI \cite{Fert1991}. For the particular case of DyCo$_x$ a few studies exist that indicate the existence of intrinsic nocollinear magnetic ground states for both single crystal and amorphous films~\cite{Yakonthos1975,Coey1976, Chen2019}. For instance, for amorphous DyCo$_{3.4}$ it is observed that the Dy moments exhibit a sperimagnetic arrangement, whereas the Co sublattice is ferromagnetically ordered~\cite{Coey1976}. 

Here, we report on the observation of skyrmions in this class of materials by investigating a ferrimagnetic DyCo$_3$(50 nm)/Ta(3~nm) film by means of x-ray magnetic circular dichroism (XMCD), small angle x-ray resonant magnetic scattering (SAXRMS), high-resolution scanning transmission X-ray microscopy (STXM) with circular polarized x-rays, and Hall transport technique. The skyrmion phase is observed for magnetic fields in the transition region to a single domain state, providing an additional topological Hall resistance that is linked to the topology of the skyrmion's magnetization texture \cite{Neubauer2013}. The magnetic microstructure of the skyrmions is found to consist of a broad shell and a point-like core, referred to as compact skyrmions~\cite{Ezawa2011, Fert2017, Caretta2018, Bernand-Mantel2020}, which can be explained by the intrinsic magnetic material constants,  thereby providing access to the strength of a DMI-like effect of the DyCo$_3$ film. 

\begin{figure}[t]
\centering
\includegraphics[width=0.92\linewidth]{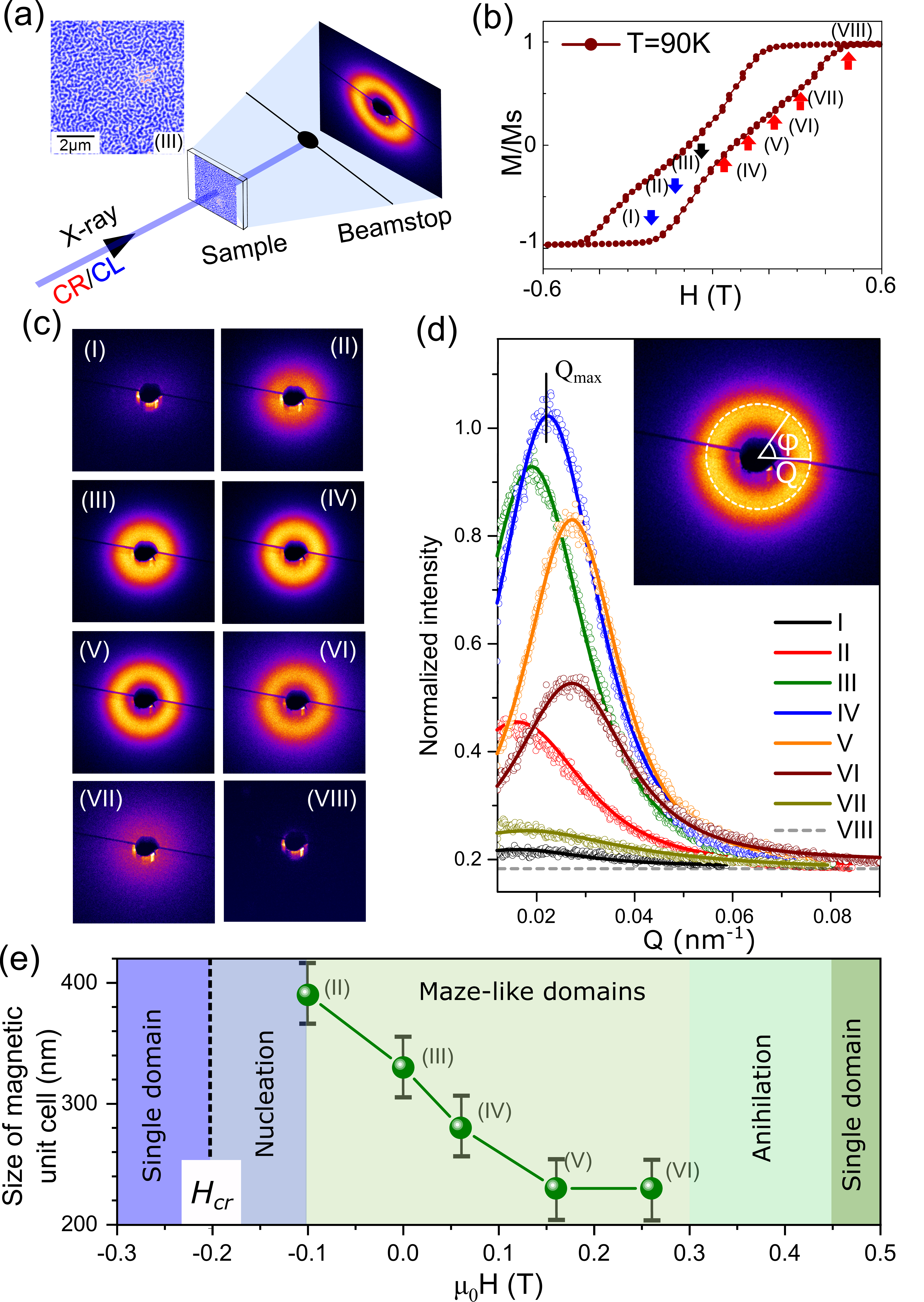}
\caption{{X-ray magnetic scattering results for DyCo$_3$ at 90~K. (a) Layout of the SAXRMS experiments: the maze-like domains produce donut-shaped scattering pattern recorded by a charge-coupled device (CCD) camera. (b) Perpendicular magnetic hysteresis loop of DyCo$_3$ obtained by XMCD. (I)-(VII) mark the fields for which the scattering patterns in (c) were required. (d) Intensity profiles extracted from the diffraction pattern via azimuthal averaging of the scattering patterns. The solid lines are fits to a split Pearson type VII distribution to determine Q$_{max}$. (e) Field dependent sizes of the magnetic unit cell of DyCo$_3$.}}
 \label{fig:1}
\end{figure}

The influence of a perpendicular magnetic field on the domain structure of DyCo$_3$ film was investigated at 90~K by means of SAXRMS. Using a beam diameter of $120\times50~\mu m^2$, this study confirms that the statements made in the following, in particular in the context of skyrmion formation, are valid over the whole macroscopic sample. In order to avoid overexposure of the CCD detector, the direct beam was blocked by a beamstop with 300~{$\mu$m} in diameter (Fig.~1a). The photon energy of the circularly polarized X-rays was fixed at 1294.3 eV corresponding to the Dy M$_5$ absorption edge, where the XMCD effect is maximized~\cite{Chen2015b} (see Fig.~2a). Fig.~1b shows the magnetic hysteresis loop obtained by XMCD at Dy M$_5$-edge and Fig.~1c shows the corresponding SAXRMS patterns for different magnetic fields labeled as (I) to (VIII) (-0.18 to 0.43 T). The sheared regions of the magnetic hysteresis loop and the corresponding donut-shaped scattering patterns as shown in Fig.~1c, reflect a maze-like domain configuration with alternating up and down domains. The radius of the ring with the maximum scattering intensity at $Q_{max}$ corresponds to the periodicity of the domain pattern, which is twice of the average domain size~\cite{Bocdanov1994, Pfau2012, Bagschik2016}: $D=\pi/Q_{max}$. Fig.~1d shows the azimuthally averaged intensity vs scattering vector Q of the patterns given in Fig.~1c. 

To determine the scattering vector Q$_{max}$ of the spectra S(Q) and hence the average domain size D as a function of magnetic field, they were fitted using a split Pearson type VII distribution \cite{Pfau2012}: 
\begin{equation}
S(Q)=S_0 \lbrack 1+\frac{(Q-Q_{max})^2}{m\alpha^2}\rbrack^{-m}
\end{equation}
The D \it {vs} \rm field behavior is given in Fig.~1e. The size of the magnetic unit cell decreases with increasing the field until the maze domain phase with a similar size of up and down domains is reached (0.05-0.26~T, states (IV)-(VI)). The average domain size (half of the magnetic unit cell) evolves from a high value equal to $\sim$200~nm right after the nucleation field region down to about $\sim$115~nm, close to the anihilation region of these magnetic textures. Note that at the border to the nucleation and anihilation regions, the reciprocal map exhibits a broad scattering pattern, which does not form a well defined donut-like shape. This is indicative for the formation of small magnetic domains.

\begin{figure}[t]
\centering
%\begin{minipage}{0.7\textwidth}
\includegraphics[width=0.95\linewidth]{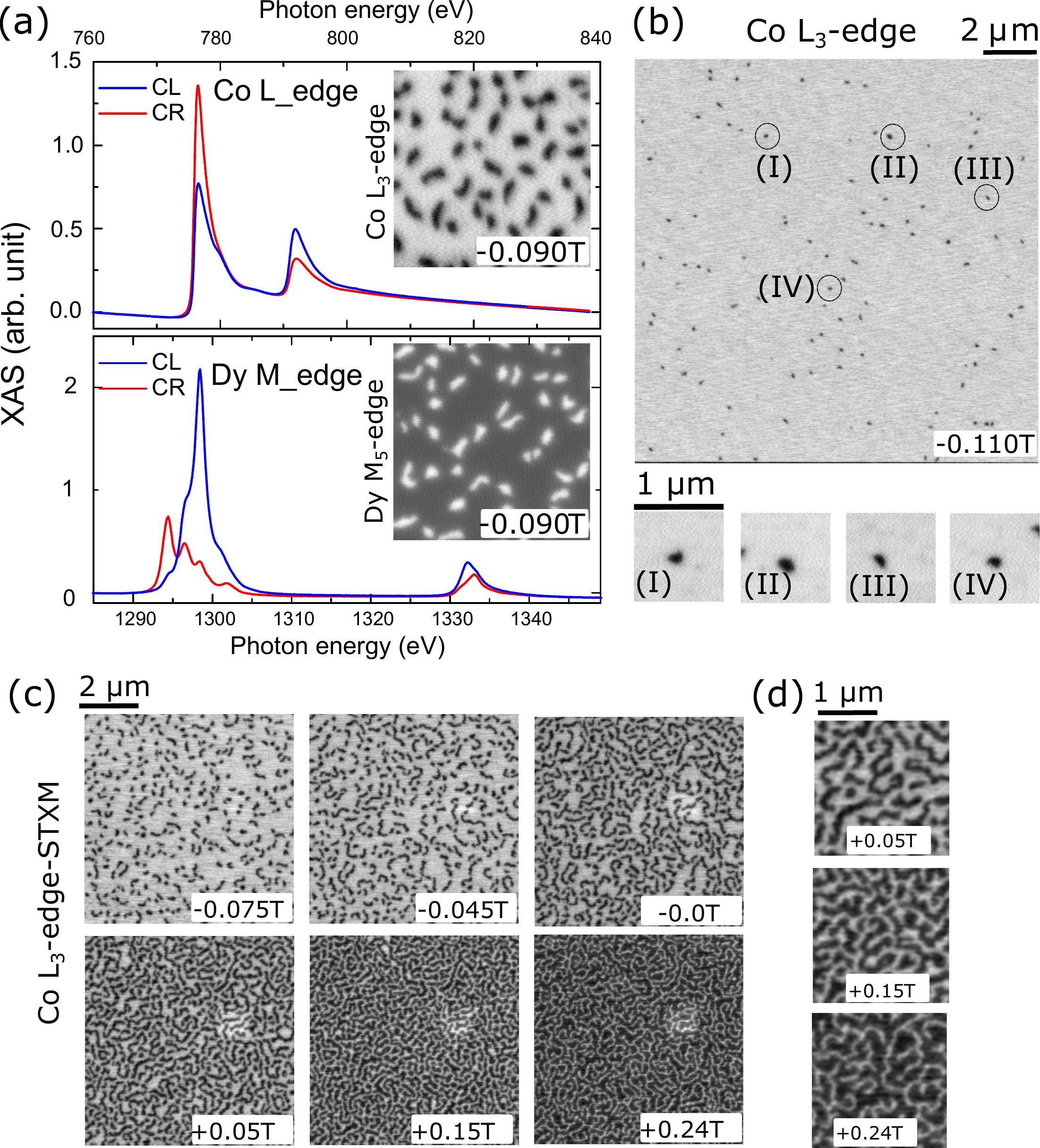}
%\end{minipage}
%\begin{minipage}{0.39\textwidth}
\caption{{{STXM of domain structures at 90~K. (a) The X-ray absorption spectra at Co L$_{2,3}$ and Dy M$_{4,5}$ edges acquired for CL and CR polarization and negative field saturation, showing the ferrimagnetic alignment of Dy and Co moments. The insets show STXM images for CL polarization of ferrimagnetic skyrmions at a field of -0.09 T with opposite contrast for Co and Dy sites. (b),(c) STXM images acquired at the Co L$_3$ edge and different perpendicular magnetic fields as labelled, displaying ferrimagentic skyrmions and maze-like domain structures (d) and the bottom panel in (b) show corresponding high resolution images.}}}
%\end{minipage}
 \label{fig:2}
\end{figure}

To reveal the details of the nature of the magnetic domain structures, in particular the eventual formation of ferrimagnetic skyrmions in this system, we performed element-specific STXM in the presence of an external perpendicular magnetic field at 90 K. The results are summarized in Figure 2. The polarized x-ray absorption spectra (XAS) at the Co L$_{2,3}$ and Dy M$_{4,5}$ edges, which were acquired at negative field saturation with circularly left (CL, blue curve) and right (CR, red curve) polarized x-rays, are shown in Fig.~2a. The XMCD signal, negative for Dy at the Dy-M$_{5}$ edge and positive for Co at the Co-L$_{3}$ edge, revealing the ferrimagnetic alignment between the Dy and Co moments, with the magnetization of Dy oriented along the magnetic field~\cite{Chen2015b}. The insets of Fig. 2a show STXM images of the skyrmions for both the Co L$_3$ and Dy M$_5$ edge. At the field of $\mu_0H_z=-0.09~T$, ferrimagnetic skyrmions are observed for both Co-L$_3$ and Dy-M$_5$ edges. The inverted contrast particularly reveals the ferrimagnetic nature of the skyrmions.

Starting from magnetic saturation the nucleation of multiple isolated ferrimagnetic skyrmions is observed at a field of -0.110~T as shown in Fig.~2b. Some individual skyrmions with quasi-circular shape are shown with higher magnification at the bottom of panel (b), demonstrating skyrmion dimensions in the range of 100~nm. When the external magnetic field is further increased, the ferrimagnetic skyrmions begin to merge, transforming into a maze-like domain structure up to +0.24~T,  as depicted in Fig. 2c, d. The latter images demonstrate the growth of Co down domains at the expense of up domains with increasing field strength.

\begin{figure}[t]
\centering
\includegraphics[width=0.92\linewidth]{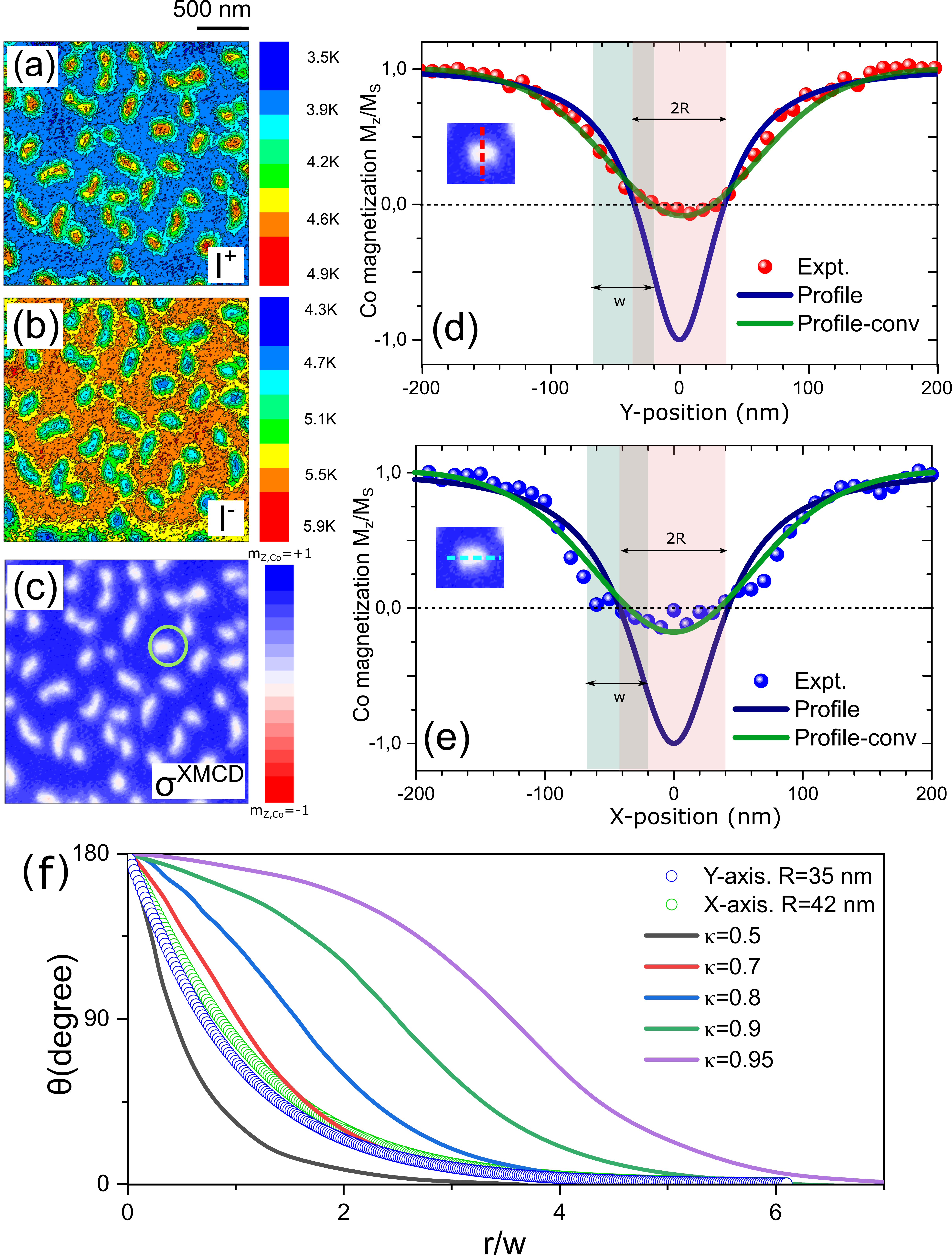}
\caption{{{Magnetic profile of ferrimagnetic skyrmions. High resolution STXM images showing the transmitted X-ray intensity $I^+$ and $I^-$ for CL (a) and CR polarized x-rays (b), respectively, resonantly tuned to the Co L$_3$-edge at 90K and $\mu_0H_{z}$=-0.09T. (c) Map of the the magnetic contrast , i.e., perpendicular component of Co magnetization $m_{z,Co} = M_{z,Co}/M_{S,Co}$ ($M_{S,Co}$: Co saturation magnetization), obtained from the XMCD intensity given as $\rm ln(I^+/I^-)$.  The rather broad skyrmion wall width $w$ and the narrow skyrmion radius $R$ result in a loss of the magnetic contrast in the skyrmion center. Vertical (d) and horizontal (e) magnetic profile of the ferrimagnetic skyrmion picked from (c) as indicated by the green circle. The dots give the experimental profiles while the green curves show the skyrmion profile according to Eq. (2) which is obtained by a deconvolution of the experimental with the Gaussian beam profile (FWHM of 65 nm). The green curves are the calculated convolutions of the beam profile and the skyrmion profile perfectly describing the experimental data. (f) Cross sections of the selected skyrmion (green and blue dots) shown together with calculated profiles (solid lines) of isolated magnetic skyrmions for various $\kappa$ taken from Ref~\cite{Bocdanov1994}.} }}
\label{fig:3}
\end{figure}

In order to quantify the lateral skyrmion profile in great detail, we performed high resolution STXM for CL and CR polarized x-rays (Fig. 3a) resonantly tuned to the Co L$_3$-edge for an external field of $\mu_0H_z$ = -0.09 T, where isolated skyrmions are present. A significant gradient in the transmitted intensity can be clearly seen down to the center of the skyrmions, indicating the formation of compact skyrmions with the inner domain reduced to a point-like core~\cite{Fert2017}. In contrast to this, bubble skyrmions exhibit thin domain wall widths and broad skyrmion radii~\cite{Buttner2018}. For compact skyrmions the skyrmion profile can be modeled by the Walker-like domain wall solution of a standard 360$^\circ$-N\'{e}el wall:
\begin{equation}
\Theta_z (r)=2\ \rm arctan\{ \frac{sinh(R/w)}{sinh(r/w)}\}
\end{equation}
where $\Theta$, $R$, and $w$ defines the polar angle of the magnetization at position r, the skyrmion radius and magnetic wall width, respectively. The magnetic contrast is shown in Fig.~3c, which is a convolution of the skyrmion profile and a two dimensional Gaussian function with a FWHM of 65 nm, corresponding to the lateral resolution of the STXM measurements. Hence, the reason for the apparent vanishing magnetic contrast in the center of the skyrmions stems from the comparable size of lateral resolution and skyrmion radius.

A quasi-circular ferrimagnetic skyrmion was selected from Fig.~3c (green circle) to investigate its profile along two orthogonal directions (Fig.~3d, e). The skyrmion profiles are well reproduced by Eq.~2 with a domain wall width of $w=50\pm5$~nm and a skyrmion core or center radius of $R$=$35\pm5$ and $42\pm5$ nm for the profiles along the vertical and horizontal direction, respectively. 

According to Eq.~2, the classification of skyrmions, whether bubble-like or compact with a point-like core, is based on the R/w ratio: for R$\gg$w, bubble skyrmions exist, while for R/w $\lesssim$ 1 compact skyrmions are formed. As pointed out by B\"{u}ttner et al., the shape of the skyrmions is determined by the parameter $\kappa=\pi D_i/(4\sqrt{AK_{eff}})$, with D$_i$, A and K$_{eff}$ representing the DMI constant, exchange stiffness, and effective anisotropy, respectively~\cite{Buttner2018}. The effective anisotropy consists of magnetocrystalline and shape anisotropy. The saturation magnetization per DyCo$_3$ cluster is 2.5-3.5 $\mu_B$ below 150 K and drops to 1.0 $\mu_B$ above 200 K \cite{Chen2015b}. Thus, at temperatures below 90~K used in this study the magnetostatic interaction is rather strong providing a significant contribution to the stabilization of skyrmions. The magnetic profile of the skyrmion shown in Fig.~3d and 3e can be well described for $\kappa$ in the range of 0.5 to 0.7 (see Fig.~3f). Considering an exchange stiffness of A=6~pJ/m \cite{Chen2015} and an effective anisotropy of K$_{eff}$=$2.5\times10^4J/m^3$, the DMI constant is estimated to be in the range of 0.25-0.35~mJ/m$^2$, which is in good agreement with the value of 0.18~mJ/m$^2$ determined for a 70~nm-thick DyCo$_4$ film\cite{Chen2019}. The domain wall width can be calculated to be $w=\pi\sqrt{A/K_{eff}}=49~nm$, which is also in agreement to the value of 50~nm obtained from the skyrmion profile (Fig.~3d-e). In the presence of {\it effective} DMI the domain wall energy $\sigma_w=4\sqrt{AK_{eff}}-\pi D_i$ corresponds to 0.61~mJ/m$^2$. 

To investigate the electrical signature of the ferrimagnetic skyrmions, magnetotransport measurements were performed using a current density of $j=4\times10^6 Am^{-2}$ driven along the x-axis and the magnetic field applied perpendicular to the plane of the film, i.e., H$\parallel$z (Fig.~4a). The Hall resistivity $\rho_{xy}$ (Fig.~4e) and longitudinal resistivity $\rho_{xx}$ (Fig.~4f) are measured at T=250, 175 and 50~K. The STXM images taken at slightly different but at comparable temperatures of 90, 200, and 300K (Fig.~4b-d) show maze-like domain states in the demagnetized state whose domain size strongly increases with temperature. Hence, the domain wall density as well  the conventional positive domain wall resistance in $\rho_{xx}(H_z)$, i.e., the difference in resistivity at zero field and saturation $\Delta\rho^{DW}$, strongly increases with decreasing temperature (Fig. 4(f)).

\begin{figure}[t]
\centering
\includegraphics[width=0.95\linewidth]{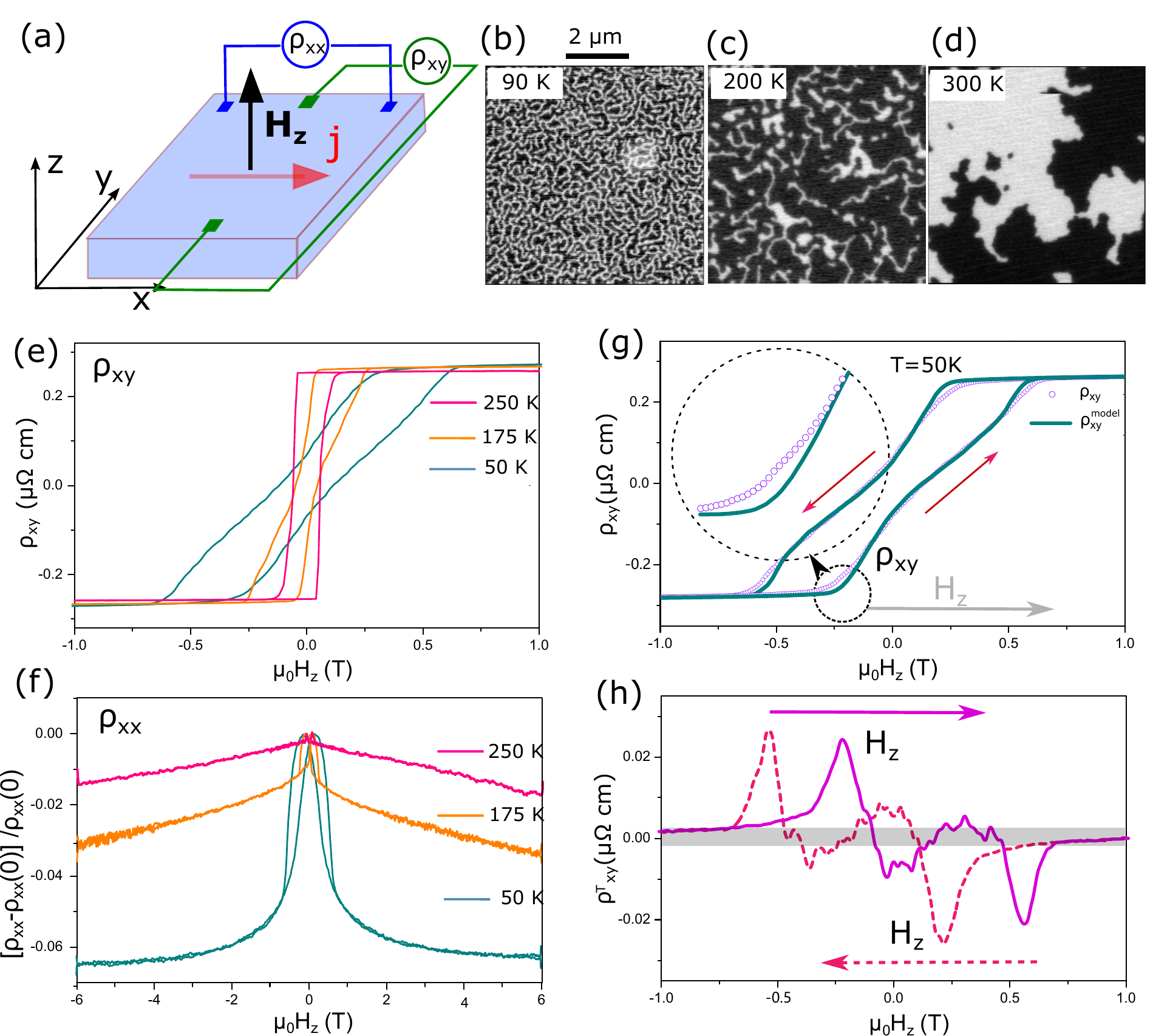}
\caption{{Topological Hall effect of DyCo$_3$. (a)Schematics for the longitudinal ($\rho_{xx}$) and Hall resistivity ($\rho_{xy}$) measurements impression using perpendicular magnetic fields (H$_z$). (e) $\rho_{xy}$ and (f) $\rho_{xx}$ vs $H_z$ behavior for 50K, 175 K, and 250 K. STXM images at T=90 (b), 200 (c), and 300K (d). (g) $\rho_{xy}(H_z)$ and $\rho_{xy}^{model}(H_z)$ at 50 K shown for positive/negative field sweep directions. The latter is obtained from the fits of the XMCD intensity. (h) Topological Hall signal $\rho^T_{xy}$=$\rho_{xy}(H_z)$-$\rho_{xy}^{model}(H_z)$, for both field sweep directions, which is non-zero in the field region where individual skyrmions in a ferrimagnetic matrix are formed. }}
 \label{fig:4}
\end{figure}

In the absence of topological entities the Hall resistivity $\rho_{xy}(H_z)$ consists of contributions from conventional and anomalous Hall effects~\cite{Nagaosa2010}:
\begin{equation}
\rm \rho^{model}_{xy}(H_z)=R_0H_z+R_SM_z(H_z)
\end{equation}
with R$_0$ the conventional Hall coefficient, and R$_S$ the cumulative anomalous Hall contribution from skew scattering, side-jump scattering, and the intrinsic (momentum space) Berry curvature mechanisms~\cite{Nagaosa2010}. 

In order to demonstrate for a topological Hall signal caused by the skyrmions~\cite{Soumyanarayanan2017}, the measured $\rho_{xy}(H_z)$ signal at 50 K (see Fig. 4e) is compared with $\rho_{xy}^{model}(H_z)$, which is obtained from the $M_z(H_z)$ behavior measured by XMCD at the same temperature according to Eq.~3, also accounting for the small normal Hall effect contribution. Obviously, in the region where individual skyrmions are present, normal and anomalous Hall effects cannot fully describe the data, therefore indicating the presence of a topological Hall effect (THE) contribution: $\rho^T_{xy}(H)=\rho_{xy}(H_z)-\rho^{model}_{xy}(H_z)$, which is related to the chirality of the local domain structures. Such a addtional signal is clearly observed as shown in Fig.~4h. The maximum value of the THE of $\sim\pm$0.03 $\mu\Omega cm$ is similar to the one observed in Ir/Fe/Co/Pt multilayers~\cite{Soumyanarayanan2017}. 

The THE has been successfully used to indicate the presence of ferromagnetic skyrmions in bulk systems and multilayers. Trivial N$\acute{e}$el or Bloch walls with chiral number of zero will not cause a THE. In ferrimagnets, however, a contribution to the THE may arise from a canting between the Co and Dy moments within the chiral domain wall region. Future investigations focusing on details of the spin structure of the domain wall will help to clarify if additional contributions to the THE in ferrimagnets exist, as speculated above.

In conclusion, we have combined soft x-ray scattering and imaging techniques with transport measurements to investigate the formation of compact ferrimagnetic skyrmions in DyCo$_3$ thin films with perpendicular magnetic anisotropy. Isolated skyrmions are imaged at 90~ K by STXM for narrow perpendicular magnetic field regions. Starting at the magnetic saturation ($m_z$= -1), the skyrmions nucleate at $\mu_0H_z$= -0.11~T continuing to merge when positively sweeping the magnetic field. A second skyrmion regime is reached when applying counter fields ($H_z>$0.25 T) that gradually annihilate the maze-like domain pattern. The detection of a topological Hall effect contribution indicates the existence of isolated skyrmions in broader field regimes at a lower temperature of 50~K. The magnetic microstructure of the skyrmions is found to consist of a small core region of 40~nm and a surprisingly broad outer wall width of 50~nm. The skyrmion profile can be reproduced when considering the intrinsic magnetic material parametes including a DMI constant of about 0.3~mJ/cm$^2$, which is consistent with previous studies on DyCo$_x$ films. The promising properties of ferrimagnetic skyrmions in DyCo$_x$ alloy thin film systems in combination with the possibility to easily tune their magnetic properties by varying their stoichiometry might be a promising route for skyrmions to be used in practical applications in future spintronic devices. In particular, the detrimental skyrmion Hall effect can be minimized when setting the compensation temperature close to room temperature, which will possibly enable the realization of functional skyrmion devices that can be operated under ambient conditions. Future systematic investigations of the skyrmion pocket in the magnetic phase diagram for DyCo$_x$ and TM-RE alloys in general will reveal their full potential.

%\section{Acknowledgments}
The authors acknowledge the financial support for the PM2-VEKMAG beamline by the German Federal Ministry for Education and Research (BMBF 05K10PC2, 05K10WR1, 05K10KE1) and by HZB. A.P.-K. gratefully acknowledges support from the DFG via Sonderforschungsbereich  (collaborative  research  center) SFB 925 (subproject B3) and S. Rudorff is acknowledged for technical support. 

\section{supplementary-methods}
%\subsection{Sample Preparation}
The samples were prepared by magnetron sputtering (MAGSSY chamber at BESSY) in an argon atmosphere of $\rm 1.5\times10^{-3}$ mbar with a base pressure of $ \rm 5\times10^{-9}$ mbar at a deposition temperature of 300~K. $\rm Si_3N_4$ membranes with a surface area of $\rm 5\times5~mm^2$ and a thickness of 100~nm were used as substrates for the soft x-ray transmission measurements including SAXRMS and STXM. A 3~nm thick Ta capping layer was grown on the $\rm DyCo_3$ layer to prevent surface oxidation. 

%\subsection{SAXRMS data analyse}
SAXRMS experiments have been performed at the VEKMAG end-station at the PM2 beamline, Helmholtz-Zentrum Berlin (HZB). The diffracted x-rays are collected on a Peltier-cooled square-shaped CCD detector covering 2.1$^\circ$ at the working distance of this study. The SAXRMS spectra (Fig.~1d) were retrieved by azimuthal averaging of the 2D patterns (Fig.~1c) after background subtraction and masking of beamstop shadow and charge scattering streaks from the membrane edges. All intensities were normalized to the charge-scattering signal from the membrane edges. The magnetic spectra were fitted with a split Pearson type VII distribution. 

%\subsection{STXM measurements}
Element-specific STXM measurements were performed at MAXYMUS, beamline UE46 at HZB in the presence of an external magnetic field, H$_z$ parallel or antiparallel to the x-ray beam. The difference between the effective spotsize (65 nm) to the ideal resultion of the used optics (32nm) can be explained by the partial coherent illumination and defocussing issues due to thermal drifts of the cryostat. The contrast of STXM image depends on the relative orientation of the sample magnetization and the X-ray polarization vector, being maximum for parallel and minimum for antiparallel alignment.  \\
%\end{methods}

\end{document}